\documentclass[useAMS,usenatbib]{mn2e}
\usepackage{amsmath,amssymb}
\usepackage{graphicx}

\title[Search for VHE Gamma Rays from 4C +55.17 with the MAGIC telescopes]{Search for Very-High-Energy Gamma Rays from the $z = 0.896$ Quasar 4C +55.17 with the MAGIC telescopes}

\author[J.~Aleksi\'c~et~al.]{
J.~Aleksi\'c$^{1}$,
S.~Ansoldi$^{2}$,
L.~A.~Antonelli$^{3}$,
P.~Antoranz$^{4}$,
A.~Babic$^{5}$,
P.~Bangale$^{6}$, \newauthor 
U.~Barres de Almeida$^{6}$,
J.~A.~Barrio$^{7}$,
J.~Becerra Gonz\'alez$^{8}$,
W.~Bednarek$^{9}$, \newauthor 
K.~Berger$^{8}$
% ...
(The MAGIC Collaboration)\thanks{Corresponding authors: 
H. Takami, email: takami@post.kek.jp, 
J. Sitarek, email: jsitarek@ifae.es, 
A. Dom\'{\i}nguez, email: albertod@ucr.edu, 
D. Paneque, email: dpaneque@mppmu.mpg.de} \\
(Affiliations can be found after the references)
}

\begin{document}

\date{Submitted \today}

\pagerange{\pageref{firstpage}--\pageref{lastpage}} \pubyear{2013}

\maketitle

\label{firstpage}

\begin{abstract}
The bright gamma-ray quasar 4C +55.17 is a distant source ($z = 0.896$) with a hard spectrum at GeV energies as observed by the Large Area Telescope (LAT) on board the {{\it Fermi}} satellite. This source is identified as a good source candidate for very-high-energy (VHE; $> 30$ GeV) gamma rays. In general VHE gamma rays from distant sources provide an unique opportunity to study the extragalactic background light (EBL) and underlying astrophysics. The flux intensity of this source in the VHE range is investigated. Then, constraints on the EBL are derived from the attenuation of gamma-ray photons coming from the distant blazar. We searched for a gamma-ray signal from this object using the 35-hour observations taken by the MAGIC telescopes between November 2010 and January 2011. No significant VHE gamma-ray signal was detected. We computed the upper limits of the integrated gamma-ray flux at 95\% confidence level of $9.4 \times 10^{-12}$ cm$^{-2}$ s$^{-1}$ and $2.5 \times 10^{-12}$ cm$^{-2}$ s$^{-1}$ above $100$ GeV and $200$ GeV, respectively. The differential upper limits in four energy bins in the range from $80$ GeV to $500$ GeV are also derived. The upper limits are consistent with the attenuation predicted by low-flux EBL models on the assumption of a simple power-law spectrum extrapolated from LAT data.
\end{abstract}


\section{Introduction} \label{sec:int}

Gamma-ray astronomy has rapidly grown in recent years thanks to the development of Imaging Atmospheric Cherenkov Telescopes (IACTs), which are able to observe gamma rays at very high energies (VHE; $> 30$ GeV), such as the High Energy gamma-ray Spectroscopic System \citep[H.E.S.S.;][]{Aharonian2006AA457p899}, the Major Atmospheric Gamma-ray Imaging Cherenkov \citep[MAGIC;][]{Aleksic2012APh35p435} telescopes, and the Very Energetic Radiation Imaging Telescope Array System \citep[VERITAS;][]{Nepomuk2009ICRCpubID1408}. Space telescopes observing in the GeV energy band such as the Large Area Telescope (LAT) on board the {\it Fermi} satellite \citep{Atwood2009ApJ697p1071} have also substantially contributed to progress in the field.

VHE photons, with energies above a characteristic energy, are strongly attenuated through electron-positron pair creation in the extragalactic background light (EBL). This characteristic EBL absorption energy is defined by the photon energy at which the pair creation optical depth is unity. The characteristic energy depends on the source's redshift. The attenuation results in a much softer spectrum at Earth than that of the intrinsic spectrum of a source \citep[e.g.,][]{Nikishov1962JETP14p393,Gould1966PRL16p252,Stecker1992ApJ390L49}. The characteristic energy is lower for sources with higher redshift. Thus, the detection of VHE gamma rays from distant sources is a challenge for IACTs. On the other hand, this absorption process brings a unique opportunity to study the EBL \citep[e.g.,][]{Stecker1993ApJ415L71,Stanev1998ApJ494L159}, which is difficult to be measured directly due to strong Galactic foregrounds \citep[e.g.,][]{Hauser2001ARAA39p249,Dwek2013APh43p112}. The EBL is mainly interpreted as stellar and dust emission integrated over the cosmic history. Therefore, VHE gamma-ray observations can be used to extract information on the imprinted cosmic stellar and galaxy evolution.

Observations of VHE gamma-ray sources with different redshifts allow us to investigate the EBL in different wavelengths and hence to examine underlying astrophysics and cosmology at different epochs. In particular, VHE gamma-ray observations of BL Lac objects at $z \lesssim 0.2$ have constrained EBL models from optical to infrared wavelengths \citep[e.g.,][]{Aharonian2006Nature440p1018,Aharonian2007AA475L9}. The most distant object ever detected in the VHE energy range, with firmly confirmed redshift, is the flat spectrum radio quasar (FSRQ) 3C 279 \citep[$z = 0.536$;][]{Albert2008Sci320p1752}. This source has also been utilized to constrain EBL models on the assumption that the intrinsic power-law index in the VHE range is softer than $-1.5$. Recently, two sources with possibly higher redshifts have been detected, namely KUV 00311-1938 \citep{Becherini2012AIPC1505p490} with $z \geq 0.506$ \citep{Pita2012AIPC1505p566} and PKS 1424+240 with $z \geq 0.6035$ \citep{Acciari2010ApJ708L100,Furniss2013ApJ768L31}, but they have not been used to estimate the EBL density due to the uncertainty in their redshift measurements. Moreover, different authors have recently claimed the detection of the EBL imprint by the statistical analyses of the spectra of blazars \citep{Ackermann2012Sci338p1190,Abramowski2013A&A550A4,Dominguez2013ApJ770p77}. In particular, the analysis of \citet{Ackermann2012Sci338p1190} covers the wide redshift range ($0.2 \lesssim z \leq 1.6$) using LAT data. \citet{Abramowski2013A&A550A4} focus on lower redshift blazars detected by H.E.S.S. ($z < 0.2$). \citet{Dominguez2013ApJ770p77} use the multi-wavelength data of the spectral energy distribution (SED) of blazars to estimate the cosmic gamma-ray horizon from the local Universe to $z \sim 0.5$ by using an EBL-model-independent technique. Although the LAT has unveiled the EBL at high redshifts ($z \sim 1$), the detection of sources with high redshifts in the VHE range allows us to test different (longer than in the LAT energy range) wavelengths of the EBL which have not yet been investigated. The MAGIC telescopes have achieved an analysis energy threshold below 100 GeV, and therefore are an ideally suited instrument among existing IACTs to study high-redshift sources.

In this paper we report the observational results of the distant quasar 4C +55.17 ($z = 0.896$) by the MAGIC telescopes. This source was recognized as a promising high-redshift candidate for VHE emission from the first LAT AGN Catalog \citep[1LAC;][]{Abdo2010ApJ715p429}. Moreover, 4C + 55.17 was also identified as a good source candidate for VHE detection by the {\it Fermi} LAT collaboration in a dedicated high-energy data analysis (private communication, September 2010), and by \citet{Neronov2011AA529A59}. The spectrum in the LAT energy range is hard and consistent with a power-law $dN / dE \propto E^{-\alpha}$ with $\alpha = 2.05 \pm 0.03$ in the range from $100$ MeV to $100$ GeV, not showing significant flux variability in the 1LAC \citep{Abdo2010ApJ715p429}. The spectral hardness of gamma-ray sources is particularly important to study the EBL because of the limited sensitivities of IACTs; harder sources can provide more photons with energies above the characteristic EBL absorption energy, allowing for studies with better precision. Even with a spectral break at the energy of several GeV recently reported \citep{Ackermann2011ApJ743p171,McConville2011ApJ738p148,Neronov2012arXiv1207.1962} this source is still a good candidate for a distant VHE gamma-ray emitter.

This source is a radio-loud active galaxy with the firmly confirmed redshift of $z = 0.896$ estimated from broad optical emission lines in its spectrum \citep{Wills1995ApJ447p139,Adelman-McCarthy2008ApJS175p297}. It had been categorized as a FSRQ, but interestingly the classification has recently been questioned due to its morphological and spectroscopic properties which are distinct from those of other FSRQs \citep{Rossetti2005AA434p449}. FSRQs are characterized by a centrally concentrated radio core with high brightness temperature and highly variable flux. However, the brightness temperature of this source is $2 \times 10^{8}$ K at 5 GHz \citep{Taylor2007ApJ671p1355}, which is much lower than that of the other known quasar-hosted gamma-ray blazars \citep{McConville2011ApJ738p148}, and a light curve in the LAT energy range is consistent with steady emission \citep{Furniss2013arXiv1303.1103}. Note that only strong variability can be significantly detected due to the limited number of detected photons by the LAT. Radio morphology has indicated that 4C +55.17 belongs to a family of young radio galaxies, that is, compact symmetric objects \citep[see][for a review]{O'Dea:1998mc}, which are a smaller version of classical Fanaroff-Riley (FR) II radio galaxies \citep{Rossetti2005AA434p449}. The low flux variability of this source also supports this indication. Importantly, this type of compact objects is a promising source candidate of ultra-high-energy cosmic rays \citep{Takami2011APh34p749} similarly to the lobes of FR II radio galaxies \citep[e.g.,][]{Biermann1987ApJ322p643,Takahara1990PTP83p1071,Rachen1993AA272p161}. \citet{McConville2011ApJ738p148} model the multi-wavelength SED of this source by using both compact symmetric object and FSRQ interpretations. The authors suggest that infrared and hard X-ray observations can help to distinguish between the two models, but VHE gamma rays also provide a clue for the distinction despite the significant EBL absorption. Also, a hadronic interpretation predicts a hard intrinsic spectrum in the VHE range \citep{Kino2011MNRAS412L20}. Therefore, observations of this source in the VHE range can provide an important clue to its nature and radiation mechanism.

This paper is laid out as follows. We describe the details of the MAGIC observations and data analysis in Section \ref{sec:obs}. Then, results are shown in Section \ref{sec:res}. Implications from the results are discussed in Section \ref{sec:dis} and a summary is presented in Section \ref{sec:sum}.

\section{Observations and Analysis} \label{sec:obs}

The MAGIC stereoscopic system consists of two IACTs each with a mirror dish diameter of 17 m, located at 2200 m above sea level at the Roque de los Muchachos, La Palma in Canary Island ($28.75^{\circ}$N, $17.86^{\circ}$W). The MAGIC telescopes have been operating in stereoscopic mode since the end of 2009, which provided an analysis energy threshold below $100$ GeV and an integral sensitivity of $0.76 \pm 0.03 \%$ of the Crab nebula flux above 300 GeV for 50-hour observations \citep{Aleksic2012APh35p435}.

4C +55.17 was observed from November 2010 to January 2011 for 21 dark nights and a total of 35 hours in stereoscopic mode. The data were taken at medium zenith angles (from $27^{\circ}$ to $37^{\circ}$) to achieve an energy threshold below $100$ GeV. The observations were performed in the so-called wobble mode \citep{Fomin1994APh2p137}; i.e. the target source direction has the offset of $0.4^{\circ}$ from the camera centre, allowing for taking both signal and background data simultaneously. The observations were performed with two pointing directions; slightly off-sources at the same declination as the source but at an angular distance of $\pm 0.4$ degree in right ascension. The direction of the wobble offset is inverted every 20 minutes to minimize systematic errors originating from possible exposure inhomogeneities.

Data were selected based on the rate of background events being in a regular range for MAGIC observations. The data selection yielded $27.73$ hours of effective time. Those data were analyzed following the standard procedure \citep{Aleksic2012APh35p435} with the MAGIC Analysis and Reconstruction Software \citep[MARS;][]{Moralejo2009arXiv0907.0943}. The analysis cuts to extract gamma-ray signals from the hadronic background were optimized independently of these observational data by means of data from Crab nebula and dedicated Monte-Carlo simulations of gamma-ray induced showers.

\section{Results} \label{sec:res}

An initial check of the energy threshold, for these observations, were made using Monte-Carlo simulations of gamma-ray induced showers. We selected Monte-Carlo events simulated under an assumed spectrum following the telescope responses after application of the same experimental cuts that were applied to the data, and made the histogram of the events as a function of input energies. The energy at the maximum of the histogram is defined as the analysis threshold. This is the standard definition of the analysis threshold for IACTs.

We assume the incident gamma-ray spectrum to be a power-law with index $\alpha = 4$. This is expected by the LAT spectrum and plausible EBL models at around 100 GeV. Note that the spectral index of the LAT spectrum above $2$ GeV (up to $60$ GeV) is $\alpha = 2.2$ in \citet{McConville2011ApJ738p148} and the spectrum above $10$ GeV can be characterized with a power-law function of $\alpha = 2.43 \pm 0.18$ \citep{FermiLAT2013arXiv1306.6772}. As a result, the energy threshold was found to be $\sim 100$ GeV. A gamma-ray signal from 4C +55.17 was searched for following the standard method using the so-called $\theta^2$ distribution, i.e., a distribution of the squared angular distance between the reconstructed arrival directions of the events and the source nominal position \citep{Daum1997APh8p1}.

\begin{figure}
\includegraphics[width=0.95\linewidth,clip]{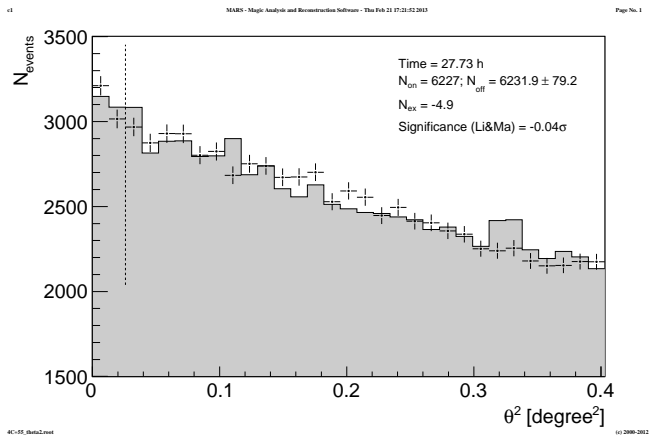}
\caption{Theta-squared distributions of events computed with respect to the positions of 4C +55.17 ({\it data points}) and the anti-source, i.e., one off-position, used for background estimation ({\it shaded region}) computed from 27.73 hours (effective time) of the MAGIC stereo observations. 
}
\label{fig:theta2}
\end{figure}

\begin{figure}
\includegraphics[width=0.95\linewidth,clip]{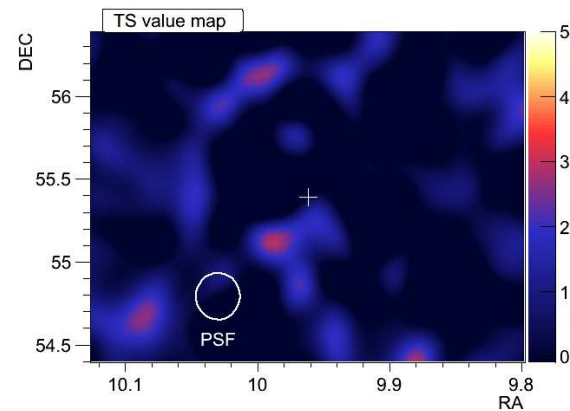}
\caption{Significance skymap of 4C +55.17 from the data corresponding to those used in Figure \ref{fig:theta2}. There is no significant signal in the direction of 4C +55.17 ({\it cross}).}
\label{fig:skymap}
\end{figure}

The $\theta^2$ distributions of the observed data with the corresponding energy threshold of $100$ GeV (i.e., images with the total number of reconstructed photoelectrons $\gtrsim 50$ in each telescope), shown in Figure \ref{fig:theta2}, indicate no significant excess of gamma rays compared to background in the direction of 4C +55.17. The significance estimated using Eq. 17 of \citet{Li1983ApJ272p317} is close to zero. Here, a region with $\theta^2 < 0.026$ deg$^2$ is used to estimate the significance. This choice is standard for $\theta^2$ cut for low-energy ($\lesssim 100$ GeV) sources. The corresponding significance map is shown in Figure \ref{fig:skymap}. It also confirms that there is no significant signal from 4C +55.17.

The upper limits on the gamma-ray flux are calculated by adding a cut in estimated photon energy to the data. The upper limits of integrated gamma-ray flux were estimated on the assumption of $\alpha = 4$. The integral upper limit above $100$ GeV is $9.4 \times 10^{-12}$ cm$^{-2}$ s$^{-1}$ at 95\% confidence level. A flux upper limit above $200$ GeV is also calculated as $2.5 \times 10^{-12}$ cm$^{-2}$ s$^{-1}$ at 95\% confidence level for comparison with a recent upper limit by VERITAS \citep[$2.6 \times 10^{-12}$ cm$^{-2}$ s$^{-1}$ for $E > 200$ GeV; ][]{Errando2011ICRC8p131}. The MAGIC upper limit is slightly lower than the upper limit of VERITAS due to the longer observation time; the VERITAS observations were performed for 17.7 hours. Further observations with VERITAS (for a total observation time of 45 hours, Furniss \& McConville 2013), resulting in a more constraining upper limit above $150$ GeV. Thanks to the lower energy threshold of MAGIC here we report an upper limit at lower energies.

We calculate differential upper limits in the 4 energy bins of equal width in logarithmic scale in the energy range between $80$ GeV and $500$ GeV. The spectral indices of the gamma-ray distributions were assumed to be $\alpha = 4$ in the first bin and $\alpha = 5$ in the other following bins because of a sharp cutoff at $\sim 100$ GeV predicted from recent EBL models (e.g., Franceschini, Rodighiero, \& Vaccari 2008; Dom\'{i}nguez et al. 2011; Gilmore et al. 2012; Inoue et al. 2013). As discussed above, the analysis threshold for IACTs is not a sharp threshold, and thus we can report on the energy range $> 80$ GeV. The results are tabulated in Table \ref{tab:difful} and plotted in Figure \ref{fig:ebl} at characteristic energies.

The upper limits on the flux depend on the assumed spectral indices for gamma-ray spectra. In order to evaluate the uncertainty of the upper limits due to this dependence, we calculate the upper limits on the assumption of harder spectra by one unit, i.e., the spectral index of $\alpha = 3$ for integral upper limits, and $\alpha = 3$ in the first bin and $\alpha = 4$ in the following bins for differential limits. As a result, the integral upper limits change by less than $10\%$ for both $> 100$ GeV and $> 200$ GeV. On the other hand, the resultant differential upper limits change within $\sim 30\%$ in the three high-energy bins and by about a factor of two in the first bin. The latter is because of a rapid change of collection area as a function of gamma-ray energy. The collection area of IACTs depends strongly on the primary gamma-ray energy. The average number of Cherenkov photons is smaller for lower energy gamma rays, making it less probable to trigger IACTs even if the shower falls within the light pool. This results in a very steep collection area around the energy threshold, and therefore strong dependence on the collection area in a finite energy bin on the assumed spectral shape.

\begin{table*}
\begin{center}
\caption{Differential upper limits of the flux and the numbers of events in on- and off-positions}
\label{tab:difful}
\begin{tabular}{ccccccc}
\hline
Range [GeV] & U.L. [cm$^{-2}$ s$^{-1}$] & U.L. [erg cm$^{-2}$ s$^{-1}$] & $\alpha$ assumed & $N_{\rm on}$ & $N_{\rm off}$ & Sign. of excess \\
\hline \hline
$79$ - $126$ & $1.0 \times 10^{-10}$ & $1.3 \times 10^{-11}$ & 4 & 6874 & 6711 & $1.4 \sigma$ \\
$126$ - $200$ & $1.6 \times 10^{-11}$ & $3.0 \times 10^{-12}$ & 5 & 1846 & 1847 & $0.0 \sigma$ \\
$200$ - $316$ & $5.7 \times 10^{-11}$ & $2.0 \times 10^{-12}$ & 5 & 714 & 706 & $0.2 \sigma$ \\
$316$ - $501$ & $3.8 \times 10^{-12}$ & $2.1 \times 10^{-12}$ & 5 & 311 & 293 & $0.7 \sigma$ \\
\hline
\end{tabular}
\end{center}
\end{table*}

\section{Discussions} \label{sec:dis}

\begin{figure}
\includegraphics[width=0.95\linewidth,clip]{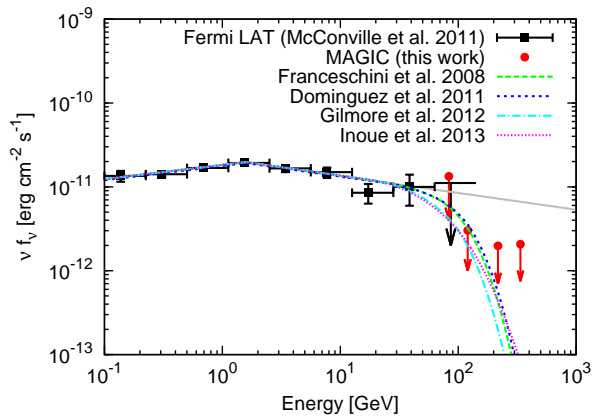}
\caption{Upper limits of differential flux at 95 \% C.L. obtained in this work ({\it circle, red}). Locations of the arrows correspond to characteristic energies in the energy ranges where the differential upper limits are calculated. The LAT spectrum derived by \citet{McConville2011ApJ738p148} {\it square, black} and its spectral fit with a broken power-law function (assumed to be the intrinsic spectrum) is depicted with black square points and the solid gray line. The intrinsic spectrum attenuated with four state-of-the-art EBL models is also shown; \citet{Franceschini2008AA487p837} with a long-dashed green curve; \citet{Dominguez2011MNRAS410p2556} with a dashed blue curve, \citet{Gilmore2012MNRAS422p3189} with a dash-dotted light-blue curve, and \citet{Inoue2013ApJ768p197} with a dotted magenta curve.}
\label{fig:ebl}
\end{figure}

We interpret the derived upper limits of the gamma-ray flux with 'low-flux' EBL models \citep{Kneiske2004AA413p807,Franceschini2008AA487p837,Dominguez2011MNRAS410p2556,Gilmore2012MNRAS422p3189,Inoue2013ApJ768p197}. Note that 'low-flux' means that the flux in the EBL models are close to that given by galaxy counts \citep[e.g.,][]{Madau2000MNRAS312L9,Fazio2004ApJS154p39,Keenan2010ApJ723p40}.

A recent study shows that the LAT spectrum of 4C +55.17 up to $\sim 60$ GeV can be well fit by a broken power-law function with a break at $\sim 2$ GeV and $\alpha = 2.2$ above the break \citep[][see Figure \ref{fig:ebl}]{McConville2011ApJ738p148}. We adopt this result and assume simple spectral extension beyond the characteristic EBL absorption energy ($\sim 100$ GeV) up to 1 TeV. This can be considered to give a plausible upper limit of the intrinsic spectrum unless another new component dominates above $60$ GeV. The assumed maximum energy up to which the source emits gamma rays is not important for our analysis because secondary photons created in electromagnetic cascades between the source and the observer will not contribute to the gamma-ray flux significantly due to the intrinsic spectral index softer than 2.

Figure \ref{fig:ebl} shows our differential upper limits and the fitted spectra 
corrected for EBL attenuation. The latter were calculated according to four low-flux EBL models \citep{Franceschini2008AA487p837,Dominguez2011MNRAS410p2556,Gilmore2012MNRAS422p3189,Inoue2013ApJ768p197}. The upper limits are close to the attenuated power-law spectra and interestingly the upper limit at $\sim 120$ GeV is slightly below the predictions of \citet{Franceschini2008AA487p837} and \citet{Dominguez2011MNRAS410p2556}. This strong upper limit was possible thanks to the low energy threshold of the MAGIC telescopes. However, we note that the flux should be regarded as consistent with the upper limit because of the dependence of the chosen binning and the statistical errors involved in the spectral fit of the LAT data. The integral upper limits above $100$ GeV and $200$ GeV provides similar results. Thus, the power-law extension is consistent with the MAGIC upper limits under the low-flux EBL models (and also higher-flux EBL models not shown). Given the redshift of 4C +55.17 ($z = 0.896$), the energies of our upper limits ($\sim 100$ GeV), and the properties of the pair production interaction, the gamma-ray photons that we are discussing are mainly attenuated by EBL photons in the ultraviolet range.

The SED of 4C +55.17 can be modeled by both an external Compton model for blazars \citep[e.g.,][]{Dermer1993ApJ416p458,Sikora1994ApJ421p153} and emission from compact radio lobes \citep{Stawarz2008ApJ680p911}. Spectral modeling in \citet{McConville2011ApJ738p148} indicates that the compact lobe model predicts a power-law spectrum above several GeV up to several hundred GeV, while the blazar model has a sharp spectral steepening at several GeV. Thus, the simple power-law intrinsic spectrum extended above $100$ GeV is an optimistic model. Since the simple power-law spectrum is allowed even for the low-flux EBL models, both models are still consistent with the upper limits. Stronger upper limits or detection in the future will allow us to test the compact symmetric object model under the low-flux EBL models.

\section{Summary and conclusion} \label{sec:sum}

The MAGIC telescopes observed 4C +55.17 for $35$ hours ($27.73$ hours of effective time with good quality data) from November 2010 to January 2011 in a wobble mode. No significant gamma-ray signal was found above $100$ GeV. Instead, both integral and differential upper limits of gamma-ray flux were derived. The integral upper limits are $9.4 \times 10^{-12}$ cm$^{-2}$ s$^{-1}$ above $100$ GeV and $2.5 \times 10^{-12}$ cm$^{-2}$ s$^{-1}$ above $200$ GeV. The differential upper limits are tabulated in Table \ref{tab:difful}. The derived limits are close to and consistent with the power-law spectrum extended from the LAT energy range attenuated by low-flux EBL models.

\citet{McConville2011ApJ738p148} claimed that a 50-hour observation with MAGIC would suffice to detect this source. However, the strict upper limits from $\sim 30$ hours of observation do not confirm this prediction, due to the medium zenith angle of the observations ($\gtrsim 27^{\circ}$), which increases the analysis threshold and suppress the performance at lowest energies \citep{Aleksic2012APh35p435}. In fact, according to our estimates, MAGIC would require $\sim 100$ hours in order to have a $5 \sigma$ detection from 4C +55.17 under the optimistic assumption of the power-law extension adopted in this study and relatively optically-thin EBL models such as \citet{Franceschini2008AA487p837} and \citet{Dominguez2011MNRAS410p2556}. After the observation of 4C +55.17, the camera of the MAGIC I telescope was upgraded, which improved the sensitivity for the lowest energies \citep{Sitarek2013arXiv1308.0141}. A new trigger system for stereoscopic observations, so-called sum-trigger \citep{Haefner2011ICRC9p246}, will improve the sensitivity below $100$ GeV significantly. These may help detect this source in the VHE range. Detailed studies on its spectrum beyond detection would require a sensitivity several times better than current ground-based gamma-ray instruments, and hence it may be a task for the next generation of instruments such as the Cherenkov Telescope Array \citep[CTA;][]{Actis2011ExA32p193,Acharya2013APh43p3}.

\section*{Acknowledgments}

We would like to thank the Instituto de Astrof\'{\i}sica de Canarias for the excellent working conditions at the Observatorio del Roque de los Muchachos in La Palma. The support of the German BMBF and MPG, the Italian INFN, the Swiss National Fund SNF, and the Spanish MICINN is gratefully acknowledged. This work was also supported by the CPAN CSD2007-00042 and MultiDark CSD2009-00064 projects of the Spanish Consolider-Ingenio 2010 programme, by grant 127740 of the Academy of Finland, by the DFG Cluster of Excellence 'Origin and Structure of the Universe', by the DFG Collaborative Research Centers SFB823/C4 and SFB876/C3, and by the Polish MNiSzW grant 745/N-HESS-MAGIC/2010/0.

\vspace*{0.5cm}

\noindent
$^{1}$ {IFAE, Edifici Cn., Campus UAB, E-08193 Bellaterra, Spain} \\
$^{2}$ {Universit\`a di Udine, and INFN Trieste, I-33100 Udine, Italy} \\
$^{3}$ {INAF National Institute for Astrophysics, I-00136 Rome, Italy} \\
$^{4}$ {Universit\`a  di Siena, and INFN Pisa, I-53100 Siena, Italy} \\
$^{5}$ {Croatian MAGIC Consortium, Rudjer Boskovic Institute, University of Rijeka and University of Split, HR-10000 Zagreb, Croatia} \\
$^{6}$ {Max-Planck-Institut f\"ur Physik, D-80805 M\"unchen, Germany} \\
$^{7}$ {Universidad Complutense, E-28040 Madrid, Spain} \\
$^{8}$ {Inst. de Astrof\'isica de Canarias, E-38200 La Laguna, Tenerife, Spain} \\
$^{9}$ {University of \L\'od\'z, PL-90236 \L\'od\'z, Poland} \\
$^{10}$ {Deutsches Elektronen-Synchrotron (DESY), D-15738 Zeuthen, Germany} \\
$^{11}$ {ETH Zurich, CH-8093 Zurich, Switzerland} \\
$^{12}$ {Universit\"at W\"urzburg, D-97074 W\"urzburg, Germany} \\
$^{13}$ {Centro de Investigaciones Energ\'eticas, Medioambientales y Tecnol\'ogicas, E-28040 Madrid, Spain} \\
$^{14}$ {Universit\`a di Padova and INFN, I-35131 Padova, Italy} \\
$^{15}$ {Technische Universit\"at Dortmund, D-44221 Dortmund, Germany} \\
$^{16}$ {Inst. de Astrof\'isica de Andaluc\'ia (CSIC), E-18080 Granada, Spain} \\
$^{17}$ {Universit\`a dell'Insubria, Como, I-22100 Como, Italy} \\
$^{18}$ {Unitat de F\'isica de les Radiacions, Departament de F\'isica, and CERES-IEEC, Universitat Aut\`onoma de Barcelona, E-08193 Bellaterra, Spain} \\
$^{19}$ {Institut de Ci\`encies de l'Espai (IEEC-CSIC), E-08193 Bellaterra, Spain} \\
$^{20}$ {Japanese MAGIC Consortium, Division of Physics and Astronomy, Kyoto University, Japan} \\
$^{21}$ {Finnish MAGIC Consortium, Tuorla Observatory, University of Turku and Department of Physics, University of Oulu, Finland} \\
$^{22}$ {Inst. for Nucl. Research and Nucl. Energy, BG-1784 Sofia, Bulgaria} \\
$^{23}$ {Universitat de Barcelona (ICC, IEEC-UB), E-08028 Barcelona, Spain} \\
$^{24}$ {Universit\`a di Pisa, and INFN Pisa, I-56126 Pisa, Italy} \\
$^{25}$ {now at Ecole polytechnique f\'ed\'erale de Lausanne (EPFL), Lausanne, Switzerland} \\
$^{26}$ {now at Department of Physics \& Astronomy, UC Riverside, CA 92521, USA} \\
$^{27}$ {now at Finnish Centre for Astronomy with ESO (FINCA), Turku, Finland} \\
$^{28}$ {also at INAF-Trieste} \\
$^{29}$ {also at Instituto de Fisica Teorica, UAM/CSIC, E-28049 Madrid, Spain} \\
$^{30}$ {now at: Stockholm University, Oskar Klein Centre for Cosmoparticle Physics, SE-106 91 Stockholm, Sweden} \\
$^{31}$ {now at GRAPPA Institute, University of Amsterdam, 1098XH Amsterdam, Netherlands}

\label{lastpage}

\end{document}